\documentclass[a4paper,11pt]{article}
\pdfoutput=1 

\usepackage{jinstpub} 

\title{\boldmath Characterization of tetrafluoropropene-based gas mixtures for the Resistive Plate Chambers of the ALICE muon spectrometer}

\author[a,1]{A. Bianchi,\note{Corresponding author.}}
\author[a]{S. Delsanto,}
\author[b]{P. Dupieux,}
\author[a]{A. Ferretti,}
\author[a]{M. Gagliardi,}
\author[b]{B. Joly,}
\author[b]{S. P. Manen,}
\author[c]{M. Marchisone,}
\author[a]{L. Micheletti,}
\author[a]{A. Rosano}
\author[a]{and E. Vercellin}

\affiliation[a]{Universit\`a degli Studi di Torino and INFN, Sezione di Torino, Via Pietro Giuria 1, 10125, Torino, Italy}
\affiliation[b]{Clermont Universit\'e, Universit\'e Blaise Pascal, CNRS/IN2P3, Laboratoire de Physique Corpusculaire, BP 10448, F-63000 Clermont-Ferrand, France}
\affiliation[c]{Institut de Physique Nucl\'eaire de Lyon, Universit\'e Claude Bernard, 4 rue Enrico Fermi, 69622, Villeurbanne, France}

\emailAdd{antonio.bianchi@unito.it}

\abstract{
The muon identification system of the ALICE experiment at the CERN LHC is based on Resistive Plate Chamber (RPC) detectors. These RPCs are operated in the so-called maxi-avalanche mode with a gas mixture made of tetrafluoroethane (C\textsubscript{2}H\textsubscript{2}F\textsubscript{4}), sulfur hexafluoride (SF\textsubscript{6}) and isobutane (\textit{i-}C\textsubscript{4}H\textsubscript{10}). All of these components are greenhouse gases: in particular, the first two gases are already phasing out of production, due to recent European Union regulations, and their cost is progressively increasing. Therefore, finding a new eco-friendly gas mixture has become extremely important in order to reduce the impact of the RPC operation on the environment, and for economic reasons.
\\
Due to the similar chemical structure, hydrofluoroolefins appear appropriate candidates to replace C\textsubscript{2}H\textsubscript{2}F\textsubscript{4} thanks to their very low GWPs, especially tetrafluoropropene (C\textsubscript{3}H\textsubscript{2}F\textsubscript{4}) with the trade name HFO1234ze.
\\
In order to identify an eco-friendly gas mixture fulfilling the requirements for operation in the ALICE environment in the coming years, a dedicated experimental set-up has been built to carry out R\&D studies on promising gas mixtures. \\
Measurements have been performed with a small-size RPC equipped with the front-end electronics, providing signal amplification, developed for ALICE operation at high luminosity after the LHC Long Shutdown 2. HFO1234ze-based mixtures with the addition of CO\textsubscript{2} are discussed in this paper as well as the role of \textit{i-}C\textsubscript{4}H\textsubscript{10} and SF\textsubscript{6} as quenchers in such mixtures.
}


\begin{document}
\maketitle
\flushbottom

\section{Introduction}
\label{sec:intro}
ALICE (A Large Ion Collider Experiment \cite{a}) is a general-purpose heavy-ion experiment at the Large Hadron Collider (LHC), studying ultra-relativistic heavy ion collisions to detect and characterize the Quark-Gluon Plasma (QGP) \cite{citQGP}.

Heavy-flavored hadrons (containing charm or beauty quarks) and quarkonia ($c\bar{c}$ and $b\bar{b}$ bound states) are sensitive probes of the medium created in ultra-relativistic heavy ion collisions \cite{citquarkonia1, citquarkonia2, citquarkonia3}. Heavy-flavors and quarkonia are studied via their muonic decay channels with the ALICE muon spectrometer \cite{a, TDR_ALICE} at forward rapidity (2.5 < $y$ < 4). The spectrometer consists of a set of absorbers, a large dipole magnet, a tracking system made of ten planes of cathode pad chambers arranged in five stations, and a muon identification system.

The muon identification system \cite{alice_trigger} of the experiment is made of 72 single-gap Resistive Plate Chambers (RPCs \cite{b}). The system has two stations, located at a distance of 16 m and 17 m from the interaction point and arranged perpendicularly around the beam pipe/shielding. Each station consists of two planes, each composed by 18 RPC detectors, covering a total area of about 5.5 $\times$ 6.5 m\textsuperscript{2} per plane. The RPC detectors have electrodes (about 270 $\times$ 70 cm\textsuperscript{2}, 2 mm thick) in phenolic bakelite with a resistivity of $10^{9} \div 10^{10}$ $\Omega$ cm \cite{c} separated by gas gap of 2 mm and the volume composition of the gas mixture is $89.7\%$ C\textsubscript{2}H\textsubscript{2}F\textsubscript{4}, $10.0\%$ \textit{i-}C\textsubscript{4}H\textsubscript{10} and $0.3\%$ SF\textsubscript{6} \cite{maxi_avalanche}. Water vapor is added to the gas mixture in order to maintain an absolute humidity of about 8000 ppm (relative humidity \textasciitilde37\%) and avoid any changes in the resistivity of the electrodes \cite{water}. The gas gap thickness uniformity is ensured by cylindrical polycarbonate spacers, placed on a $10 \times 10$ cm\textsuperscript{2} grid. The inner surface of the bakelite, which is in contact with the gas mixture, is coated with a double linseed oil layer \cite{arnaldi2003aging}, while the outer surface is coated with a conductive graphite paint. One of the graphite layers is connected to the high voltage (\textit{HV}), whereas the other one is connected to the ground, in order to apply a uniform electric field across the gas gap. 

Each RPC is read out with copper strips on both sides in order to obtain two-dimensional position information. The pitch and length of the read-out strips vary within each station to keep the occupancy of the detector constant. During the LHC Run 1 (2010-2013) and Run 2 (2015-2018), the RPCs were equipped with the ADULT front-end electronics (FEE) \cite{d}, which has no amplification and a minimum discrimination threshold of \textasciitilde 7 mV. The ADULT FEE allows one to operate the detectors in the so-called maxi-avalanche mode \cite{saturation1, saturation2, saturation3}, with the gas mixture described above and a working \textit{HV} between \textasciitilde10.0 and \textasciitilde10.4 kV \cite{martinoRPC2010}. 

The RPC detectors have shown stable operation and an efficiency greater than $98\%$ from the beginning of Run 1 up to now \cite{Ferretti_RPC2018}. During the LHC Run 3 (2021 onwards), ALICE will take data in Pb-Pb collisions at an instantaneous luminosity \textasciitilde6 times higher than in Run 2, thanks to a major upgrade of both the LHC injectors and the experiment \cite{Antonioli_TRD}. In view of the harsher running conditions, a new amplified FEE, called FEERIC \cite{e}, has been developed. This will allow one to operate the RPCs in pure avalanche mode, with a lower charge per hit (by a factor 3$\div$5), hence improving the rate capability and reducing ageing effects. One RPC of the ALICE muon system has been equipped with FEERIC prototypes since the beginning of Run 2, with fully satisfactory performance \cite{f}.

The current ALICE mixture is not environment-friendly because of the presence of greenhouse gases. In this paper we report the results of the R\&D studies on eco-friendly gas mixtures. A reasonable goal could be to operate the change of gas mixture between the LHC Run 3 and Run 4, during the LHC Long Shutdown 3 (2024--2026).

The paper is organized as follows. In Section~\ref{sec:motivation} we discuss the approach to find an alternative to C\textsubscript{2}H\textsubscript{2}F\textsubscript{4}. The experimental set-up for the R\&D studies is presented in Section~\ref{sec:setup}, whereas the results of these tests are reported in Section~\ref{sec:results}. Finally, conclusions are drawn in Section~\ref{sec:conclusions}.

\section{Search for environment-friendly mixtures}
\label{sec:motivation}

A greenhouse gas is classified according to its Global Warming Potential (GWP), which is a relative measurement of how much heat is trapped into the atmosphere with respect to the same mass of CO\textsubscript{2}.
In general fluorinated gases may have very high GWPs. In fact, the gas mixture of the ALICE muon RPCs features C\textsubscript{2}H\textsubscript{2}F\textsubscript{4} (GWP
\textasciitilde 1300) and a small concentration of SF\textsubscript{6} (GWP \textasciitilde 23500), while the GWP of \textit{i-}C\textsubscript{4}H\textsubscript{10} is only 3 \cite{IPCC}. The current ALICE mixture has an overall GWP of 1237, almost $95\%$ of which is due to the presence of C\textsubscript{2}H\textsubscript{2}F\textsubscript{4}.

Recent regulations \cite{h} from the European Union (EU) impose the reduction of the emissions of fluorinated greenhouse gases (F-gases) in EU countries since January 2015. The major EU provision is a gradual phase out of the hydrofluorocarbons (such as C\textsubscript{2}H\textsubscript{2}F\textsubscript{4}) available on the market in order to limit the total amount of the production. In any case gas mixtures for most applications must have a GWP lower than 150. Scientific activities are excluded from these restrictions, nevertheless CERN is pushing the LHC experiments to replace the greenhouse gases with eco-friendly ones. In addition, the phasing out of F-gases will lead to a progressive increase of their price due to the limited future availability.

For these reasons, R\&D studies on eco-friendly gas mixtures for the ALICE-muon RPCs are necessary.

In order to reduce the GWP of the current ALICE mixture, one of the possible approaches relies on the replacement of C\textsubscript{2}H\textsubscript{2}F\textsubscript{4} with a chemically similar molecule \cite{m, i, liberti2016}, but with a low enough GWP.

The tetrafluoropropene C\textsubscript{3}H\textsubscript{2}F\textsubscript{4} is similar to C\textsubscript{2}H\textsubscript{2}F\textsubscript{4}. The tetrafluoropropene is an olefin, so it presents a double bond in the chain of carbon atoms. The allotropic form of tetrafluoropropene HFO1234ze (trans-1,3,3,3-Tetrafluoroprop-1-ene) is the most interesting for our purposes because it is not flammable at room temperature and its GWP is lower than 1. HFO1234ze has a boiling point equal to about -19 $^{\circ}$C at atmospheric pressure. This gas is compatible with most common polymeric materials and toxicity tests are in accordance with applicable guidelines and standards, e.g. it is considered safe for refrigeration and air-conditioning applications.

We have performed R\&D studies on RPCs with HFO1234ze-based gas mixtures and the FEERIC electronics. The new mixture has to provide similar performance as the current one \cite{f}. The approach to the problem is exclusively experimental, because several parameters of HFO1234ze, such as the electron collision cross sections, the photon absorption spectrum and the chemical reactivity are still unknown. This prevents a more theoretical approach and the implementation of reliable simulations.

\section{The experimental set-up}
\label{sec:setup}
A dedicated experimental set-up has been used to carry out R\&D studies on various gas mixtures, flushing one small-size (50 $\times$ 50 cm\textsuperscript{2}, 2 mm gas gap) RPC with the same features of the RPCs used in ALICE. 

The detector is horizontally placed and exposed to the cosmic-ray flux. The applied high voltages (\textit{HV}\textsubscript{appl}) are corrected for temperature and atmospheric pressure. All the results of this work will be given in terms of the corrected high voltage: $HV = HV\textsubscript{\normalfont{appl}} \cdot \frac{p_{0}}{p} \cdot \frac{T}{T_{0}}$, where $p$ is the pressure, $T$ is the temperature, $p_{0}$ and $T_{0}$ are reference values and are equal to 1000 mbar and 293.15 K, respectively.
The cosmic-ray trigger is provided by a three-fold coincidence of plastic scintillators, coupled with photomultipliers. The trigger area is about $4 \, \times \, 8$ cm\textsuperscript{2}, while the average trigger rate is $0.12$ Hz. The read-out strips have 2 cm pitch. The strip plane on the anode side of the RPC is equipped with FEERIC front-end cards, configured to discriminate negative signals. 

FEERIC \cite{e} is designed in the AMS 350 nm CMOS technology and is able to amplify signals up to $\pm$1 pC in the linear range. Its intrinsic time resolution is lower than 1 ns (root mean square) in the input charge range 100-1000 fC, which is the typical input charge expected for the RPCs in avalanche mode.

The four adjacent strips covering the trigger area are read out on both ends: on one end the signals are discriminated by FEERIC, while on the other end the analogic signals are summed by a fan-in/fan-out module and then digitized by an oscilloscope. This allows for the simultaneous measurement of the efficiency, with FEERIC, and of the amplitude of the signals. The digital oscilloscope (LeCroy WaveSurfer 510), used for measuring the amplitude signals, has a bandwidth of 1 GHz and sampling rate of 10 GS/s.

The efficiency of the RPC detector is measured by analyzing its response in the presence of a trigger signal. If at least one strip is fired, the detector is considered efficient. The threshold used for the data acquisition is \textasciitilde 130 fC (70 mV after the amplification \cite{f}), which is the same setting used for the FEERIC prototypes installed in ALICE.

\begin{figure}[h]
\centering
\includegraphics[scale=0.57]{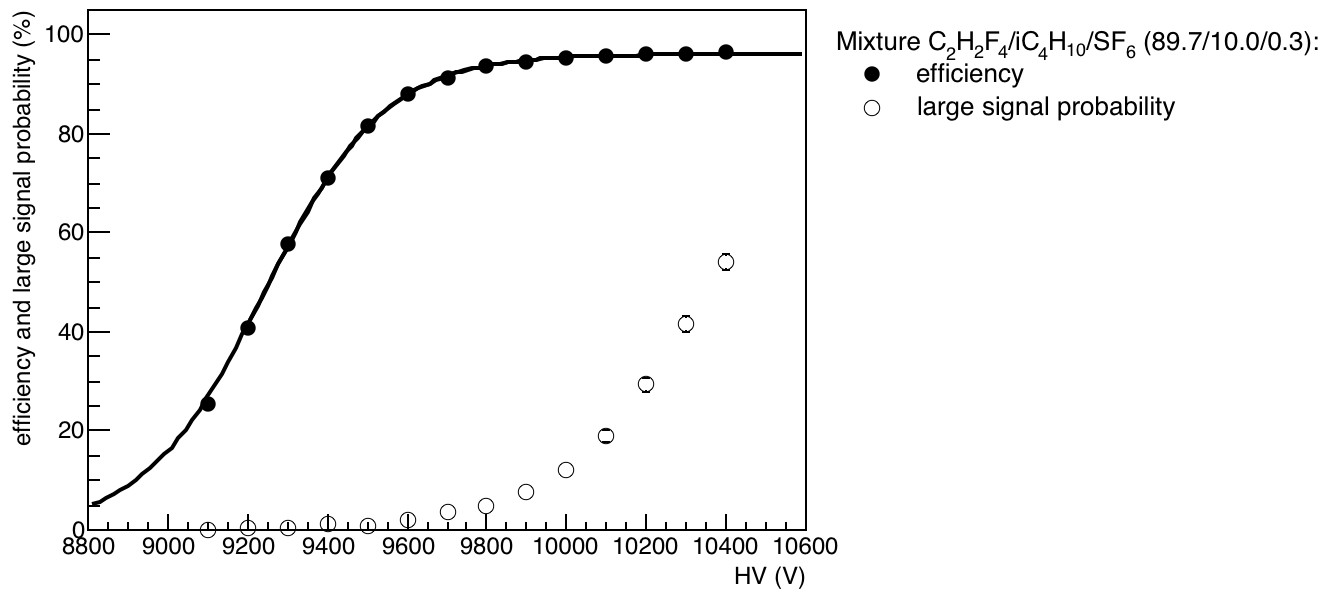}
\caption{\label{fig:2} Efficiency curve and large signal probability of a RPC detector equipped with FEERIC front-end electronics (with threshold set to 70 mV) measured with cosmic rays. The solid line is a sigmoid-function fit to the efficiency curve (see text for details).}
\end{figure}

First of all, a measurement of the efficiency curve versus \textit{HV} has been performed with the ALICE mixture in order to characterize the detector. The results\footnote{The efficiency curves as a function of \textit{HV} are fitted by the following sigmoidal function:
\begin{equation}
\epsilon(HV) = \frac{\epsilon\textsubscript{\normalfont{max}}}{1 + \exp[- \alpha \cdot (HV - HV_{0.5})]}
\end{equation}
where $\epsilon\textsubscript{\normalfont{max}}$ is the efficiency for $HV \xrightarrow{} \infty$, $HV_{0.5}$ is the $HV$ value for which the efficiency is equal to half of $\epsilon\textsubscript{\normalfont{max}}$ and $\alpha$ is the slope at the inflection point. The voltage region $\Delta HV_{\epsilon_{0.1 \xrightarrow{} 0.9}}$, where the efficiency rises from $10\%$ to $90\%$ of its maximum, can be easily calculated as $4.4 / \alpha$, thanks to the symmetry of the sigmoidal function.} are reported in figure~\ref{fig:2}. The maximum efficiency of the RPC with the ALICE mixture is \textasciitilde95\% in this specific experimental set-up.

During the charge multiplication in the gas gap, the high density of free charges and UV photons may induce the transition from an electron avalanche to a highly-saturated avalanche or streamer \cite{sauli}, which liberate more charge in the gas and induce larger signals on the electrodes with respect to a standard avalanche. The presence of large signals is not desirable because they imply a reduction in the rate capability of the detector caused by the large local voltage drop, which is slowly recovered because of the high resistivity of the electrodes. In addition, large-charge deposition events may enhance ageing effects \cite{arnaldi2003aging}. It is well known in literature (see e.g. \cite{saturation3}) that SF\textsubscript{6} plays a crucial role in reducing the avalanche size, avoiding unreasonably large signals and suppressing streamers in C$_2$H$_2$F$_4$-based mixtures.

\begin{figure}[h]
\centering
\includegraphics[scale=0.7]{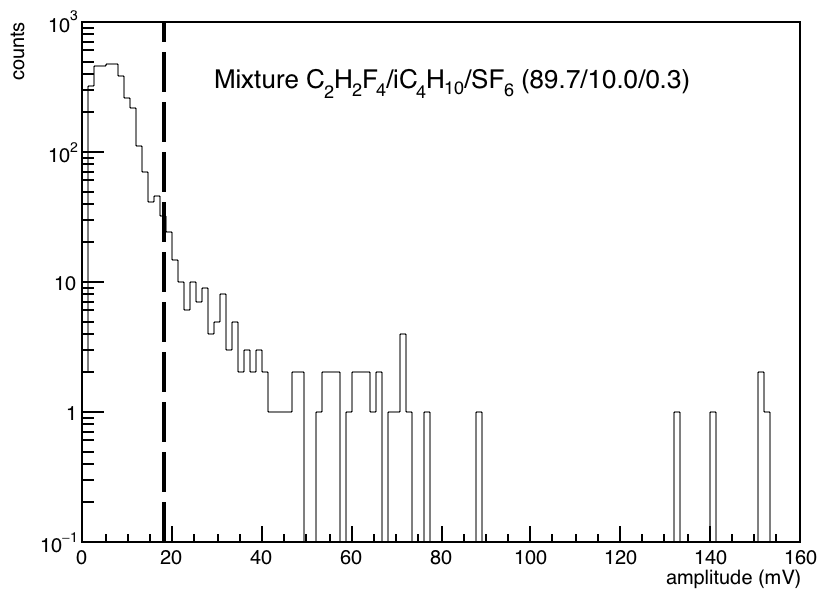}
\caption{\label{fig:3} Amplitude spectrum of 3500 signals induced in the trigger area at 9.8 kV with the ALICE mixture. The dashed line illustrates the amplitude threshold of 18 mV.}
\end{figure}

Figure~\ref{fig:3} shows the amplitude of signals induced on the strips when the RPC is operating at the working point (9.8 kV) with the ALICE mixture. The amplitude spectrum of the signals is a long-tailed distribution, as shown in figure~\ref{fig:3}, and we used an amplitude threshold of 18 mV to tag events with a large signal amplitude. In this work, the fraction of large signals is used, along with the efficiency, to assess the performance of the tested gas mixtures. In figure~\ref{fig:2} the large signal probability versus \textit{HV} is shown for the ALICE mixture. The large signal probability, evaluated with this threshold, is about 5\% at the working point with the ALICE mixture.

\section{Methodology and results}
\label{sec:results}
The direct replacement of C\textsubscript{2}H\textsubscript{2}F\textsubscript{4} with C\textsubscript{3}H\textsubscript{2}F\textsubscript{4} in the ALICE mixture is not advisable because it leads to RPC operating voltages higher than 14 kV, which would very likely result in current instabilities and high detector noise.

In order to lower the working point, we carried out many tests with addition of Ar, CO\textsubscript{2}, O\textsubscript{2} and N\textsubscript{2} in the mixture at different concentrations. CO\textsubscript{2} turned out to be the most promising solution.

In this section the results of the HFO1234ze-based mixtures with the addition of CO\textsubscript{2} are shown. All gas mixtures presented in this paper are made of four components. Since the contribution of each gas in the mixtures is complex to investigate, we proceeded by changing the fractions of two gas components out of four at a time, evaluating how their ratio affects the performance of the RPC.

\subsection{Variation of the ratio between HFO1234ze and CO\textsubscript{2}}

\begin{figure}[h]
\centering
\includegraphics[scale=0.57]{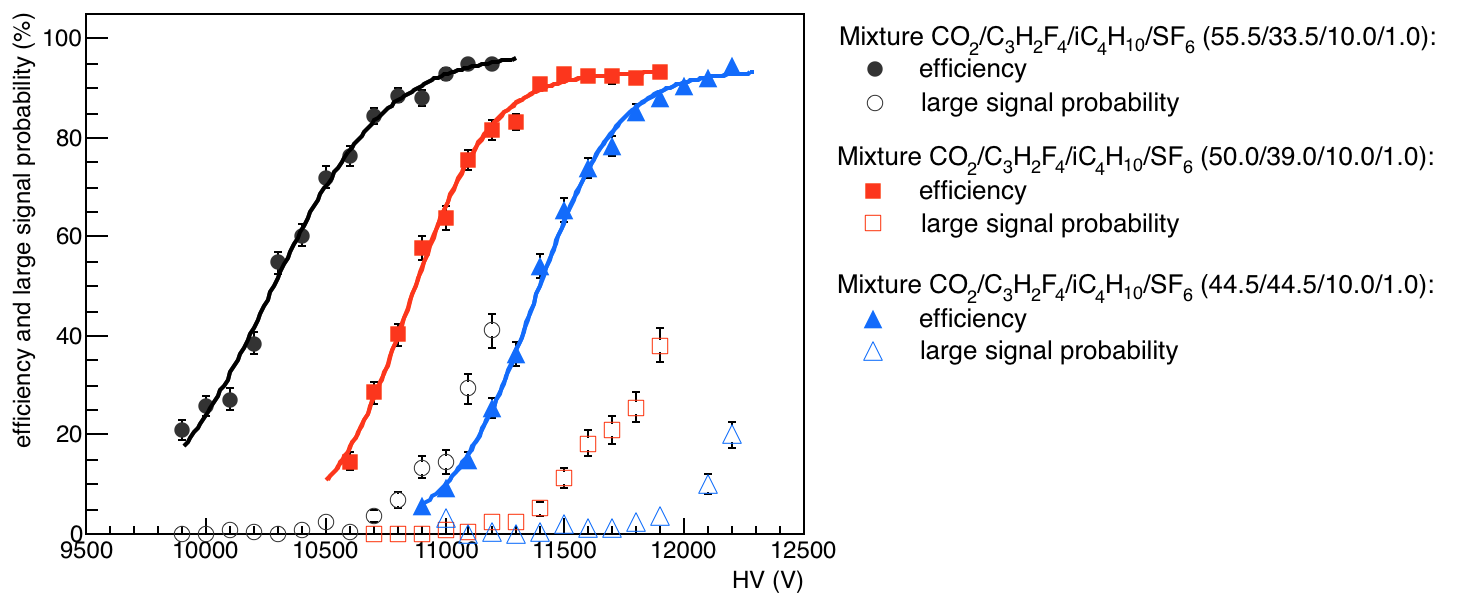}
\caption{\label{fig:4} Efficiency and large signal probability for gas mixtures with different ratios between C\textsubscript{3}H\textsubscript{2}F\textsubscript{4} (HFO1234ze) and CO\textsubscript{2}, while the concentrations of \textit{i-}C\textsubscript{4}H\textsubscript{10} and SF\textsubscript{6} are kept constant (10\% and 1\%, respectively).}
\end{figure}

In order to study the effect of HFO1234ze and CO\textsubscript{2} on the efficiency of the RPC and the fraction of large signals, we analysed three different cases: without \textit{i-}C\textsubscript{4}H\textsubscript{10}, with $10\%$ \textit{i-}C\textsubscript{4}H\textsubscript{10} and $20\%$ \textit{i-}C\textsubscript{4}H\textsubscript{10}, while the fraction of SF\textsubscript{6} is always 1\%.
If the proportion of CO\textsubscript{2} is increased and HFO1234ze is decreased, the working point turns out to be shifted towards lower voltages as shown in figure~\ref{fig:4}, where the fractions of \textit{i-}C\textsubscript{4}H\textsubscript{10} and SF\textsubscript{6} are $10\%$ and $1\%$, respectively. In the same figure, the large signal probability as a function of \textit{HV} is shown.

\begin{figure}[h]
\centering
\includegraphics[scale=0.57]{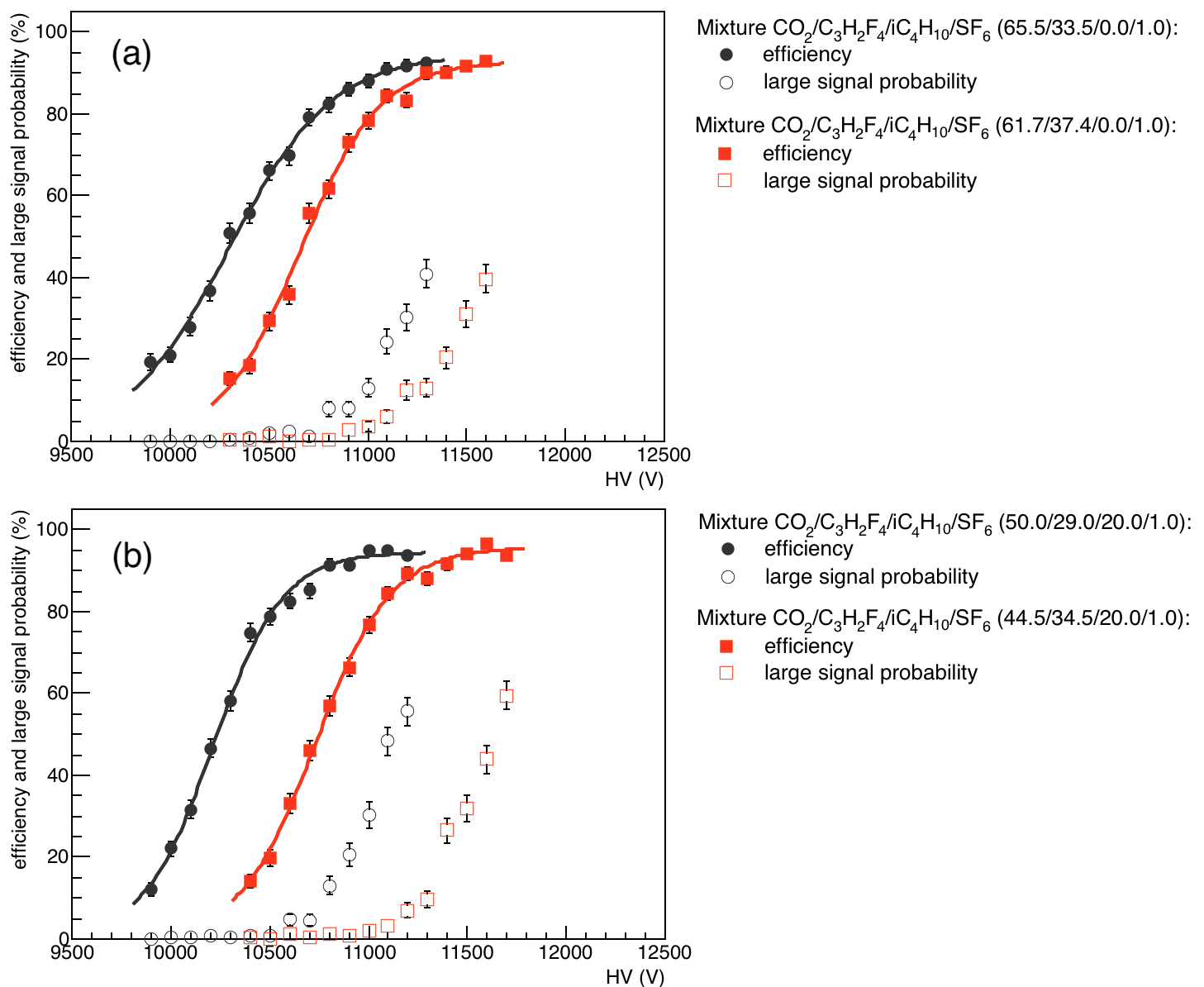}
\caption{\label{fig:4bis} Efficiency and large signal probability for gas mixtures without \textit{i-}C\textsubscript{4}H\textsubscript{10} (a) and with $20\%$ \textit{i-}C\textsubscript{4}H\textsubscript{10} (b) for different ratios between C\textsubscript{3}H\textsubscript{2}F\textsubscript{4} (HFO1234ze) and CO\textsubscript{2}, while the concentration of SF\textsubscript{6} is kept constant at 1\%.}
\end{figure}

A similar trend of the working point position as a function of the HFO1234ze fraction has been observed without \textit{i-}C\textsubscript{4}H\textsubscript{10} and with $20\%$ \textit{i-}C\textsubscript{4}H\textsubscript{10}, as reported in these figure~\ref{fig:4bis}a and~\ref{fig:4bis}b. In the same figure, the large signal probability as a function of \textit{HV} is also shown.

\subsection{Variation of the ratio between HFO1234ze and \textit{i-}C\textsubscript{4}H\textsubscript{10}}

\begin{figure}[h]
\centering
\includegraphics[scale=0.57]{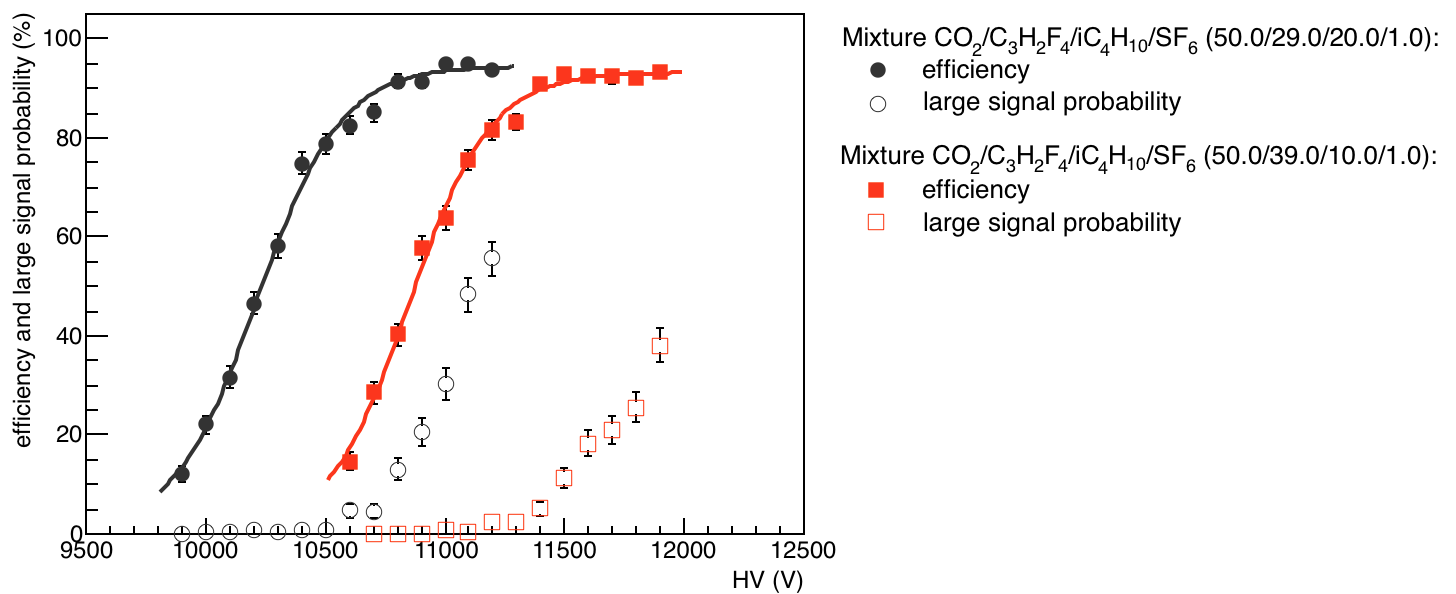}
\caption{\label{fig:6} Efficiency and large signal probability for gas mixtures with different ratios between C\textsubscript{3}H\textsubscript{2}F\textsubscript{4} (HFO1234ze) and \textit{i-}C\textsubscript{4}H\textsubscript{10}, while the concentrations of CO\textsubscript{2} and SF\textsubscript{6} are kept constant (50\% and 1\%, respectively).}
\end{figure}

In order to explore the effect of \textit{i-}C\textsubscript{4}H\textsubscript{10} in the mixture, we evaluated the RPC performance at different ratios between HFO1234ze and \textit{i-}C\textsubscript{4}H\textsubscript{10}. The working point of the detector is shifted towards lower voltages if the \textit{i-}C\textsubscript{4}H\textsubscript{10} is increased and HFO1234ze is decreased, as shown in figure~\ref{fig:6}. In the same figure, the large signal probability as a function of \textit{HV} is shown.

The correlation between the increase of the working point with the ratios HFO1234ze/CO\textsubscript{2} (figures~\ref{fig:4} and~\ref{fig:4bis}) and HFO1234ze/\textit{i-}C\textsubscript{4}H\textsubscript{10} (figure~\ref{fig:6}) might be explained by the electron attachments to HFO1234ze \cite{n}. Indeed, the effect of the electron attachment turns out to be dependent on the partial pressure of the HFO1234ze at the atmospheric pressure, as shown by the measurements of its electron swarm parameters \cite{n}. Therefore the concentration of this gas strongly affects the total gain of the gas mixture.

\subsection{Variation of the ratio between CO\textsubscript{2} and \textit{i-}C\textsubscript{4}H\textsubscript{10}}

The effect of the concentration of \textit{i-}C\textsubscript{4}H\textsubscript{10} has been carefully studied because of its impact on the gas mixture flammability. The reduction, or even better the removal, of this flammable component in the gas mixture would be desirable for safety and practical reasons. In particular, the substitution of \textit{i-}C\textsubscript{4}H\textsubscript{10} with CO\textsubscript{2} has been investigated.

\begin{figure}[h]
\centering
\includegraphics[scale=0.57]{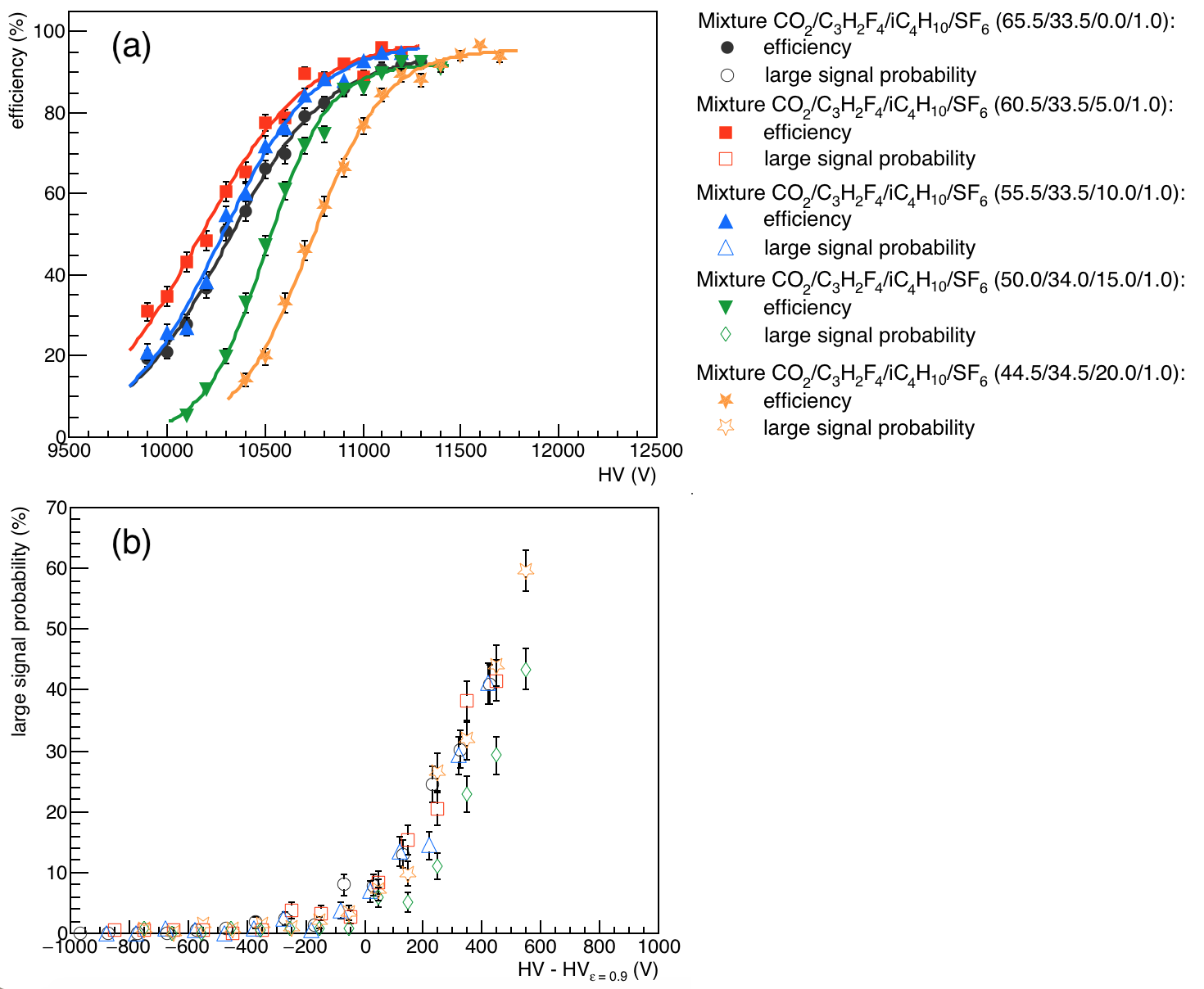}
\caption{\label{fig:8} Efficiency (a) and large signal probability (b) for gas mixtures with different ratios between \textit{i-}C\textsubscript{4}H\textsubscript{10} and CO\textsubscript{2}, while the concentrations of C\textsubscript{3}H\textsubscript{2}F\textsubscript{4} (HFO1234ze) and SF\textsubscript{6} are kept almost constant.}
\end{figure}

Figure~\ref{fig:8}a shows the efficiency curve at different ratios of \textit{i-}C\textsubscript{4}H\textsubscript{10} and CO\textsubscript{2}, while the concentrations of HFO1234ze and SF\textsubscript{6} are kept constant (about $34\%$ and $1\%$ respectively). The fraction of large signals is shown in figure~\ref{fig:8}b as a function of the $HV - HV_{\varepsilon = 0.9}$, where $HV_{\varepsilon = 0.9}$ is the voltage where the efficiency is 90\% of its maximum value ($\epsilon\textsubscript{\normalfont{max}}$). $HV_{\varepsilon = 0.9}$ is obtained, for each mixture, by fitting the data with a sigmoidal function. 
\\
The shift of the working points is not monotonic versus the ratio CO\textsubscript{2}/\textit{i-}C\textsubscript{4}H\textsubscript{10}. The fraction of large signals does not appear to be significantly reduced with the increase of \textit{i-}C\textsubscript{4}H\textsubscript{10}, as shown in figure~\ref{fig:8}, except for the fact that the large signal probability rises slightly less steeply with 15\% \textit{i-}C\textsubscript{4}H\textsubscript{10} than with lower fractions of this gas. Finally, we observe that the voltage region ($\Delta HV_{\epsilon_{0.1 \xrightarrow{} 0.9}}$), where the efficiency rises from $10\%$ to $90\%$, becomes smaller as the \textit{i-}C\textsubscript{4}H\textsubscript{10} fraction increases, as reported in table~\ref{tab:1}.
\\
The interplay of \textit{i-}C\textsubscript{4}H\textsubscript{10} and CO\textsubscript{2} in HFO1234ze-based mixtures is rather complex and needs further studies in order to explore the feasibility of RPC mixtures without flammable components.

\begin{table}[htbp]
\centering
\smallskip
\begin{tabular}{|c|c|}
\hline
Gas mixture & $\Delta HV_{\epsilon_{0.1 \xrightarrow{} 0.9}} [kV]$\\
\hline
65.5\% CO\textsubscript{2}, 33.5\% C\textsubscript{3}H\textsubscript{2}F\textsubscript{4}, 0.0\% \textit{i-}C\textsubscript{4}H\textsubscript{10}, 1.0\% SF\textsubscript{6} & $1.14 \pm 0.06$\\

60.5\% CO\textsubscript{2}, 33.5\% C\textsubscript{3}H\textsubscript{2}F\textsubscript{4}, 5.0\% \textit{i-}C\textsubscript{4}H\textsubscript{10}, 1.0\% SF\textsubscript{6} & $1.23 \pm 0.07$\\

55.5\% CO\textsubscript{2}, 33.5\% C\textsubscript{3}H\textsubscript{2}F\textsubscript{4}, 10.0\% \textit{i-}C\textsubscript{4}H\textsubscript{10}, 1.0\% SF\textsubscript{6} & $1.07 \pm 0.05$\\

50.0\% CO\textsubscript{2}, 34.0\% C\textsubscript{3}H\textsubscript{2}F\textsubscript{4}, 15.0\% \textit{i-}C\textsubscript{4}H\textsubscript{10}, 1.0\% SF\textsubscript{6} & $0.70 \pm 0.03$\\

44.5\% CO\textsubscript{2}, 34.5\% C\textsubscript{3}H\textsubscript{2}F\textsubscript{4}, 20.0\% \textit{i-}C\textsubscript{4}H\textsubscript{10}, 1.0\% SF\textsubscript{6} & $0.84 \pm 0.04$\\
\hline
\end{tabular}
\caption{\label{tab:1} Values of voltage region $\Delta HV_{\epsilon_{0.1 \xrightarrow{} 0.9}}$, where the efficiency rises from $10\%$ to $90\%$, at different ratios between CO\textsubscript{2} and \textit{i-}C\textsubscript{4}H\textsubscript{10}.}
\end{table}

\subsection{Variation of the ratio between HFO1234ze and SF\textsubscript{6}}

\begin{figure}[h]
\centering
\includegraphics[scale=0.57]{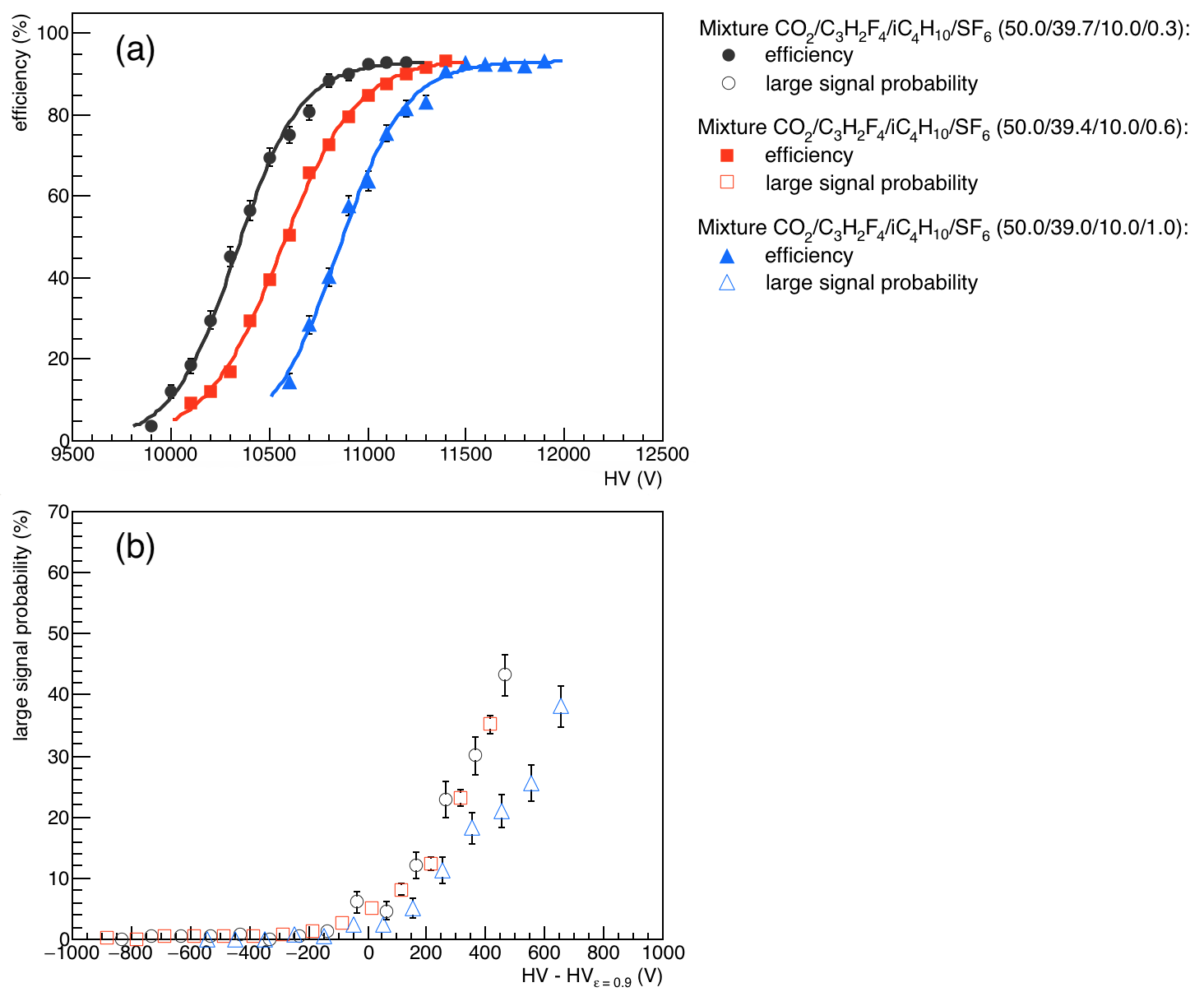}
\caption{\label{fig:5} Efficiency (a) and large signal probability (b) for gas mixtures with different concentrations of SF\textsubscript{6}.}
\end{figure}

In order to study the suppression of large signals, we evaluated how the  percentage of SF\textsubscript{6} affects the RPC performance, while the concentrations of CO\textsubscript{2} and \textit{i-}C\textsubscript{4}H\textsubscript{10} are kept constant. Figure~\ref{fig:5}a shows that the working point of the RPC detector increases with the addition of SF\textsubscript{6}. A small variation on the concentration of SF\textsubscript{6} leads to an important effect on the working point: the shift of the working point is \textasciitilde 500 V from 0.3\% to 1.0\% SF\textsubscript{6}.
In figure~\ref{fig:5}b the impact of the SF\textsubscript{6} percentage on the large signal probability is shown as a function of $HV - HV_{\varepsilon = 0.9}$.
\\
No significative changes in the large signal probability are observed if SF\textsubscript{6} is increased from 0.3\% to 0.6\%, while the suppression of large signals is slightly higher with 1.0\% SF\textsubscript{6}.

\subsection{Promising gas mixtures with low GWP}

Eco-friendly gas mixtures with similar features of the ALICE mixture might be promising to substitute the current mixture of the ALICE-muon RPCs. The parameters used to identify a promising gas mixture are the high voltage of the working point, along with the large signal probability. The working point should not be too much larger than the working point of the ALICE-muon RPCs. The fraction of large signals, which include streamers, has to be as low as possible. Of course, the GWP of the promising gas mixtures must be lower than that of the ALICE mixture.

\begin{figure}[h]
\centering
\includegraphics[scale=0.57]{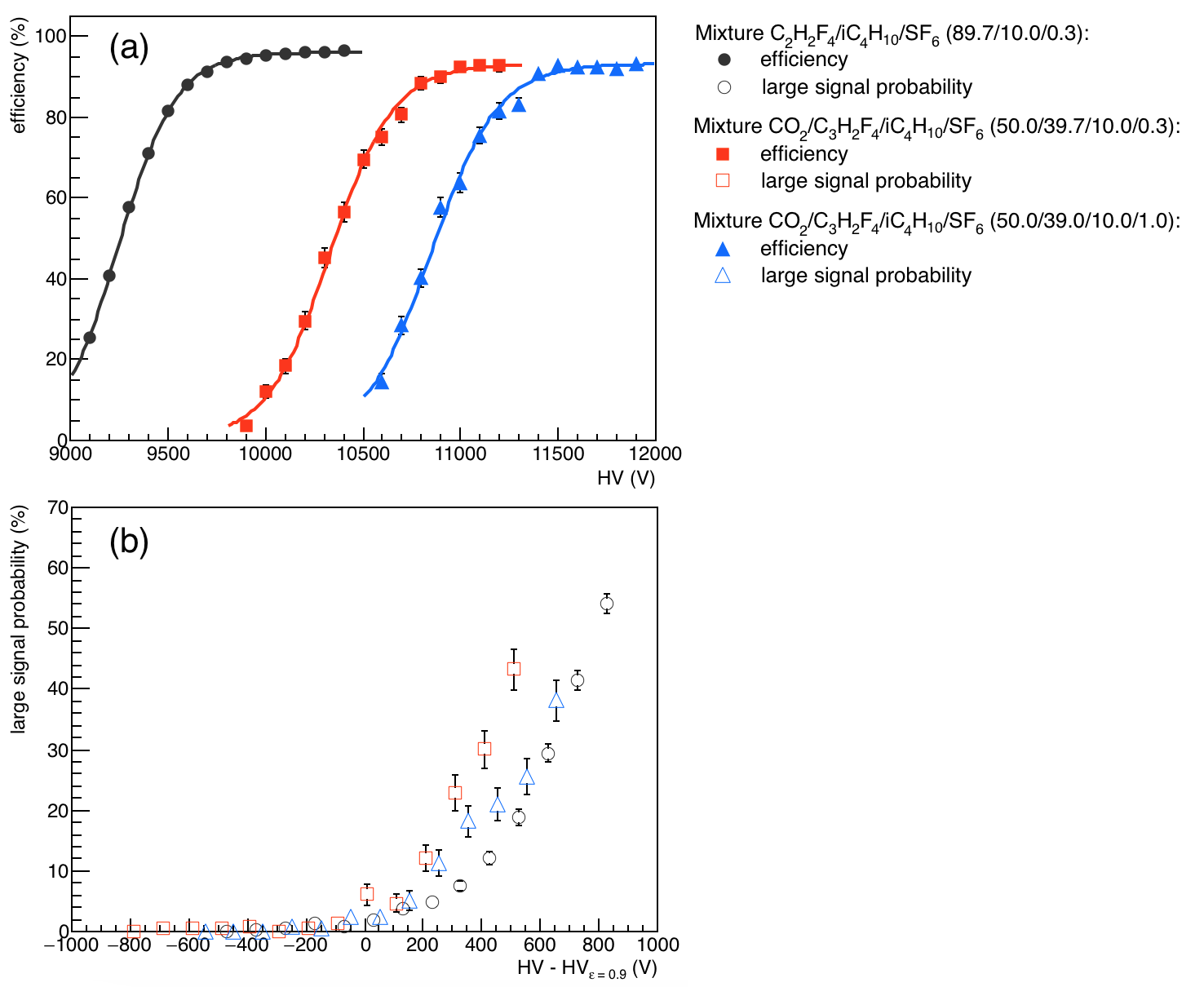}
\caption{\label{fig:9} Efficiency (a) and large signal probability (b) for the standard ALICE mixture and the most promising C\textsubscript{3}H\textsubscript{2}F\textsubscript{4}-based gas mixtures.}
\end{figure}

Figure~\ref{fig:9} shows the most promising gas mixtures, among those tested so far:
\\
$50\%$ CO\textsubscript{2}, $39.7\%$ HFO1234ze, $10\%$ \textit{i-}C\textsubscript{4}H\textsubscript{10}, $0.3\%$ SF\textsubscript{6} and $50\%$ CO\textsubscript{2}, $39\%$ HFO1234ze, $10\%$ \textit{i-}C\textsubscript{4}H\textsubscript{10}, $1\%$ SF\textsubscript{6}.
Their GWPs are respectively 71 and 236, while the GWP of the current ALICE mixture is 1237.
\\
As shown in figure~\ref{fig:9}a, the working point of the detector with the first mixture ($50\%$ CO\textsubscript{2}, $39.7\%$ HFO1234ze, $10\%$ \textit{i-}C\textsubscript{4}H\textsubscript{10}, $0.3\%$ SF\textsubscript{6}) is quite close to the working point of the ALICE RPCs during LHC Run 1 and Run 2, which should be safe for operation, while the large signal probability is not as low as in the current ALICE mixture. On the contrary, the working point with the second mixture ($50\%$ CO\textsubscript{2}, $39\%$ HFO1234ze, $10\%$ \textit{i-}C\textsubscript{4}H\textsubscript{10}, $1\%$ SF\textsubscript{6}) is \textasciitilde 500 V higher, but the suppression of large signals is higher, as reported in figure~\ref{fig:9}b.

\section{Conclusions and outlook}
\label{sec:conclusions}

Studies on HFO1234ze-based mixtures have shown a strong dependence between the working point of the detector and the concentration of HFO1234ze, most probably due to its electron attachment. Promising results are obtained with the addition of CO\textsubscript{2} in a HFO1234ze-based mixture.

SF\textsubscript{6} turns out to play a crucial role in suppressing large signals in HFO1234ze-based mixtures, as was the case for C\textsubscript{2}H\textsubscript{2}F\textsubscript{4}-based mixtures.

The increase of \textit{i-}C\textsubscript{4}H\textsubscript{10} in place of CO\textsubscript{2} does not seem to play a crucial role in reducing the large signal probability, which is promising in light of a possible future reduction or removal of this (flammable) gas from the ALICE mixture. Unfortunately, the reduction of \textit{i-}C\textsubscript{4}H\textsubscript{10} in the mixtures studies so far leads to a less sharp efficiency turn-on, which calls for more effort in this direction.

The most promising gas mixtures, which have been found so far, consists of $50\%$ CO\textsubscript{2}, $39.7\%$ HFO1234ze, $10\%$ \textit{i-}C\textsubscript{4}H\textsubscript{10}, $0.3\%$ SF\textsubscript{6} and $50\%$ CO\textsubscript{2}, $39\%$ HFO1234ze, $10\%$ \textit{i-}C\textsubscript{4}H\textsubscript{10}, $1\%$ SF\textsubscript{6}. The GWPs of these two gas mixtures are significantly lower than the GWP of the current ALICE mixture. 

Measurements of the RPC performance under irradiation, e.g. at the CERN Gamma Irradiation Facility \cite{gif}, will be crucial for assessing the rate capability and ageing properties resulting from the considered mixtures, thus qualifying these for high-luminosity operation at the LHC.


\end{document}